\documentclass[global,twocolumn]{svjourarx}
\usepackage{lmodern}	
\usepackage[utf8]{inputenc}
\usepackage[T1]{fontenc}
\usepackage{textcomp}
\usepackage{graphicx}
\usepackage{placeins}
\usepackage[labelformat=brace]{subcaption} 
\usepackage{float}
\usepackage{amsmath}

\usepackage{tikz, pgfplots}
\pgfplotsset{compat=1.17, /pgfplots/error bars/error bar style={line width=1.0pt, solid}}
\usetikzlibrary{spy}

\usepackage[square,numbers]{natbib}			
\graphicspath{{./graphics/}{./}}

\begin{document}

\title{Holographic imaging of an array of submicron light scatterers \\ at low photon numbers}
\author{Sebastian Kölle\and Manuel Jäger \and Markus Müller \and Wladimir Schoch \and Wolfgang Limmer \and Johannes Hecker Denschlag} 
\institute{Institut für Quantenmaterie, Universität Ulm, 89081 Ulm, Germany}
\titlerunning{\,\,}

\maketitle

\begin{abstract}
We experimentally test a recently proposed holographic method for imaging coherent light  
 scatterers which are distributed over a 2-dimensional grid. 
 In our setup the scatterers consist of a back-illuminated, opaque mask with submicron-sized holes. 
 We study how the imaging fidelity depends on various parameters of the set-up. 
We observe that a few hundred scattered photons per hole already suffice to obtain a fidelity of 96\%
to correctly determine whether a hole is located at a given grid point. The holographic method
demonstrated here has a high potential for applications with ultracold atoms in optical lattices.  
\end{abstract}

\section{Introduction} \label{sec:intro}

In recent years, ultracold atoms in optical lattices have become a promising platform for fundamental research of many-body and solid-state physics as well as for applications in quantum information.
Quantum gas microscopes have been developed which use fluorescence imaging to detect atomic distributions in 2D optical lattices, resolving single atoms at individual lattice sites, see e.g. \cite{Bak09,She10,Che15,Hal15,Par15,Gro15}.
   In these quantum gas microscopes an individual atom typically scatters thousands of photons. This leads to heating, and therefore additional cooling techniques 
  such as Raman sideband cooling \cite{Par15} are typically required to prevent the atom from leaving its lattice site during imaging. 
  A main motivation for the work presented here was to test a site-resolved imaging method 
  with small numbers of scattered photons so that additional cooling is not needed.  
  Indeed, fluorescence imaging with small photon numbers  and single-atom sensitivity 
  has been recently demonstrated \cite{Ber18}, but only for atoms propagating in free space.
  Furthermore,  other imaging methods for atoms exist which are not based on fluorescence imaging. 
 For example, these include spatially resolved ionization of atoms followed by ion detection \cite{Ger08,Vei21}.
 A review on various single-atom imaging techniques can be found in Ref. \cite{Ott16}. 
 Holographic imaging of cold atomic clouds has been developed and demonstrated in recent years,
 see e.g. \cite{Kad01,Tur05,Wu14}, but not yet with $\mu$m- and single-atom-resolution. 
\begin{figure}
	\includegraphics[scale=0.85]{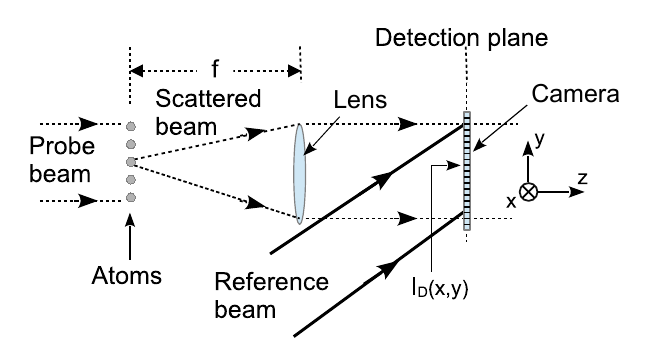}
	\caption{Coherent light scattered by the atoms is superimposed with a reference laser beam of the same frequency.  For clarity, only the scattered light of one atom is shown. The atomic array is located in the objective's front focal plane.  The digital camera sensor is the detection plane.}
	\label{fig:beams}
\end{figure}\\
We have recently proposed a novel approach to site-resolved detection  of atoms in a 2D optical lattice which is based on holographic techniques \,\cite{Hof16}.
The main idea is schematically illustrated in Fig.\,\ref{fig:beams}. An ensemble of atoms is exposed to a near-resonant laser beam from which they coherently scatter light via fluorescence. The scattered light is collimated by a lens and then superimposed with a collimated reference laser beam of the same frequency. The resulting interference pattern is recorded
  by a digital camera sensor. 
  A fast Fourier transform (FFT) of the recorded interference pattern $I_\mathrm{D}(x,y)$ yields a site-resolved image of the atomic distribution in the lattice. 
 The role of the reference beam is to amplify the weak atomic signals and to shift 
 the information on the atomic distributions in the hologram to the FFT positions where technical background noise is small. 
Our calculations predicted that this holographic imaging is better than 99\% error-free already for about 200 scattered photons per atom. 
Therefore, as a rough estimate, for a lattice which is deeper than a few times 200 photon recoil energies, holographic imaging might work without additional cooling. For example, for $^6$Li where the recoil energy is 3.5\,$\mu$K$\times k_B$ for the resonant wavelength of 671\,nm, a trap depth of about 2~mK$\times k_B$ should be sufficient to keep the atoms trapped in their respective lattice sites. Here, $k_B$ is the Boltzmann constant.
\begin{figure}
\includegraphics[scale=0.9]{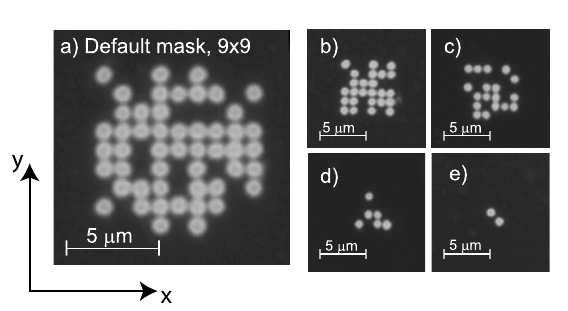}
\caption{Pictures of various hole masks taken with an optical microscope.
	Each hole is circular and has a radius of 0.3$\pm$0.03\,$\mu$m.  For each mask, the holes are positioned on a  9 $\times$ 9 square lattice with a lattice constant of 1\,$\mu$m. 
		The hole pattern within each mask is arbitrary. } 
\label{fig:masks}
\end{figure}\\
In this work, we take a first experimental step to  test our proposed holographic detection scheme.
For this, we replace the atomic scatterers  by an array of circular sub\-micron-sized holes 
in an opaque flat mask, see Fig.\,\ref{fig:masks}. 
The mask is homogeneously back-illuminated with laser light which is diffracted when passing through the holes.
The  holes are randomly arranged in a square lattice with $1 \mu$m lattice constant, similar to the distribution of real atoms in a partially occupied 2D optical lattice \cite{note1}. 
Clearly, this setup is much simpler than working with an array of cold atoms, yet it offers all necessary 
ingredients for the scheme.\\
Besides experimentally demonstrating holographic imaging, we
measure the fidelity of reconstructing the hole positions of the known mask. 
We study how this recognition fidelity depends on various parameters such as the scattered photon number,
the reference laser power and the incidence angle of the reference laser.  
We discuss various noise sources and resolution limits and we investigate how to optimize the setup given these limits. 
We find that about 200 diffracted photons per hole are sufficient to reconstruct the hole positions in the masks with a fidelity of 96\%.

\vspace{-0.3cm}
\section{Experimental setup} \label{sec:experiment}

The hole masks were fabricated in the cleanroom of the Microelectronics Technology Center, University of Ulm, via e-beam lithography. 
Details of the fabrication can be found in Appendix \ref{sec:holemasks}. 
\begin{figure}[h]
	\includegraphics[scale=0.95]{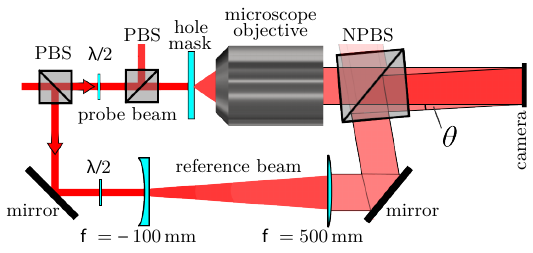}
	\caption{Scheme of the experimental setup. 
		The  671 nm laser beam (red color)  is split by a PBS into a probe beam (upper path) and a reference beam (lower path). A $\lambda/2$ plate and PBS control the intensity of the probe beam which illuminates the hole mask.		
		The diffracted light is collected  by an infinity-corrected microscope objective lens and hits the sensor of a digital camera  as a collimated beam. A tilted  NPBS is used to superimpose this probe beam light with the
		 collimated reference beam. The resulting interference pattern is recorded by the camera sensor. }
	\label{fig:scheme}
\end{figure}

A scheme of the holography setup is depicted in Fig.\,\ref{fig:scheme}. It resembles the one for digital holographic microscopy which is based on a Mach-Zehnder interferometer \cite{Kim11,Ver15}.
The beam of a laser with wavelength $\lambda$ = 671\,nm and  $\approx 1$ MHz linewidth is split by a polarizing beam splitter (PBS) into a probe beam and a reference beam. The probe beam is attenuated by the combination of a $\lambda/2$ plate and a second PBS. It is diffracted at the hole mask and the diffracted
light is collected  by an infinity-corrected microscope objective with numerical aperture (NA) between 0.5 and 0.75, 
(e.g. Zeiss Epiplan Neofluar 50x, 0.75 HD Dic, 44 23 55 with an effective focal length of about $f = 4$\,mm). 
The distance between the back side of the objective lens and the sensor is 16~cm. 
Since the distance between mask and objective equals the focal length $f$, light scattered
from a hole in the mask is collimated by the lens and subsequently  propagates as a plane wave with a beam diameter of about 5~mm towards the camera. 
The diffracted probe beam and the reference beam are merged at a tilted non-polarizing beam splitter (NPBS) such that they overlap well at the camera in the detection plane. While the diffracted probe beam hits the camera approximately under vertical incidence, the reference beam has a small tilt angle  $\theta\approx 1^{\circ}$. 
The reference beam is roughly Gaussian with a waist of  5.3\,mm  and a power of 120 $\mu$W  behind the NPBS. A cross section of the beam profile is shown in Fig.\,\ref{fig:line_profiles}, labelled as $I_R$.
The beam illuminates the CMOS sensor chip (13.3 mm $\times$ 13.3 mm) of the digital camera  
 pco.edge 4.2LT which  has 2048$\times$2048 pixels. Further details on the camera can be found in Appendix \ref{sec:digicam}.
 We verified that measurements with a broader and thus more uniform reference beam profile did not produce a higher recognition fidelity.  
  The exposure time $t_{exp}$ was typically 144\,$\mu$s and the intensity of the reference beam was set such that the linear detection range of the camera sensor was optimally used while avoiding
 saturation.  This intensity corresponds to a peak photon number per pixel of about 40,000.
  In the following, we show how the hole pattern of the mask is 
reconstructed via FFT from the holographic image taken by the digital camera.

\section{Reconstruction of the  hole pattern of the mask} 
\label{sec:recognition}

The light intensity distribution $I_\mathrm{D}(x,y)$ in the sensor plane of the digital camera is given by
\begin{align}
I_\mathrm{D}(x,y) &= \frac{c\epsilon_0}{2} \left| E_\mathrm{S}(x,y) + E_\mathrm{R}(x,y)\right|^2  \nonumber \\
&= \,\frac{c\epsilon_0}{2} (| E_\mathrm{S} |^2 + | E_\mathrm{R} |^2) + c\epsilon_0\, Re\{   E_\mathrm{S} E_\mathrm{R}^\ast  \} , 
\label{eq:ID_SR1}
\end{align}
where $E_\mathrm{S}$ and $E_\mathrm{R}$ are the electric fields  (in complex notation) of the diffracted and reference beams  in the detection plane, respectively.  $c$ is the speed of light in vacuum and $\epsilon_0$ is the permittivity of free space.
In the limit of a very weak scattered light field we can neglect  the term $| E_\mathrm{S} |^2$.
Ideally, the term $| E_\mathrm{R} |^2$ is just a constant. 
The information about the hole pattern is contained in the third term, the interference term.

For simplicity, we first consider a single hole $n$ in the mask at position $\mathbf{r}_n  = (x_n,y_n) $
 which emits a scattered, spherical light wave. The lens at the focal distance $f$ collimates the wave into a plane wave with the wavevector component $\mathbf{k}_n $ in 
$(x,y)$ direction,
\begin{equation}
	\mathbf{k}_n = 	\frac{k}{\sqrt{x_n^2 + y_n^2 + f^2}} \left( \begin{array}{c} -x_n \\ -y_n  \end{array} \right)
	\approx  -\frac{k}{f}\mathbf{r}_n
	 ,
	\label{eq:k_n}
\end{equation}
where $k = 2\pi/  \lambda$ is the wavenumber of the light. The origin of the coordinate system is located on
the optical axis of the microscope lens. The approximation in Eq. \eqref{eq:k_n} is valid for holes close to the optical axis, i.e. \mbox{$x_n, y_n \ll f$}.
At the camera sensor, this plane wave interferes with the plane wave of the reference beam
with wavevector  $\mathbf{k}_R$, leading to a 
2D sinusoidal fringe pattern $\propto \cos((\mathbf{k}_n -  \mathbf{k}_R)  \cdot \mathbf{r}  + \varphi)$. 
 Here,  $\mathbf{r}  = (x,y) $ is the position vector in the sensor plane of the camera
 and $\varphi$ is a constant phase. 
 The FFT of this pattern produces an output that only contains two single peaks at  $\pm(\mathbf{k}_n -  \mathbf{k}_R)$, corresponding to  opposite momenta. 
  After subtraction of the constant vector  $\mathbf{k}_R$ we obtain 
 $\mathbf{k}_n$ which, according to Eq. (\ref{eq:k_n}), corresponds to the hole position $\mathbf{r_n}$,
  apart from a factor $-k/f$. 
 The constant vector $\mathbf{k}_R$ depends on the incidence angle of the reference beam with respect to the detection plane. 
 In spherical coordinates  we have 
 \begin{align}
 \mathbf{k}_R = k \left( \begin{array}{c} \sin\theta \cos\phi \\ \sin\theta \sin\phi  \end{array} \right),
 \end{align}
where $\theta$ and $\phi$ are the polar and azimuthal angles of the reference beam, respectively.

If there is more than one hole in the mask, each hole contributes
a corresponding sinusoidal pattern. All these patterns  add up linearly  under the condition
that the reference beam has much higher intensity than the scattered probe beam. Since the FFT is a linear operation it reproduces the hole pattern of the mask.

\begin{center}
\begin{figure}
	\includegraphics[scale=0.9]{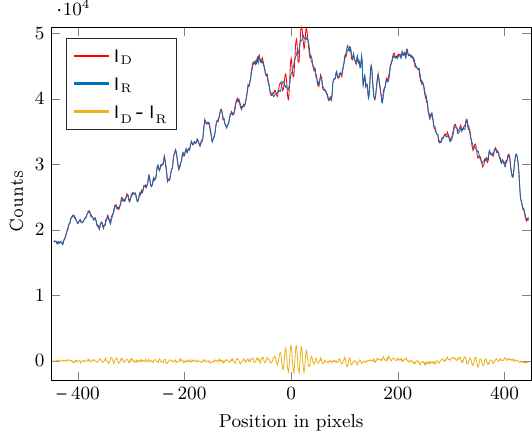}
	\caption{Line profiles  of the digital images of the 
		hologram $I_D(x,y)$ and the reference beam  $I_R(x,y) = c\epsilon_0|E_R|^2 /2$, as well as their difference  $I_{D} - I_R$.  Here, $x$ and $y$ are positions in units of pixels. The profiles are taken along the $x$-direction around the $y$-center (i.e. $y = 0$) of the image. In order to reduce noise we have averaged over 11 pixel rows, for details see \cite{note2}.}
	\label{fig:line_profiles}
\end{figure}
\end{center}

 In practice, the $| E_\mathrm{R} |^2$ term in Eq. (\ref{eq:ID_SR1}) is not just a constant, but it exhibits 
 corrugations e.g. due to diffraction from dust on top of optical surfaces.
   This hampers the  reproduction of the hole pattern. We find that
   most of these perturbations can be removed by subtracting an image $I_R$ taken with only the reference beam by blocking the probe beam ($E_\mathrm{S}=0$) and averaging over 30 recordings to reduce noise. 
 Figure \,\ref{fig:line_profiles}  shows  line profiles  of the hologram
 before (red) and after the $I_R$-subtraction (yellow). 
 The blue line is the profile of the reference signal. 
  The line profiles run along the $x$-direction through the $y$-center of the hologram.

\begin{figure}[h]
\includegraphics[width=0.48\textwidth]{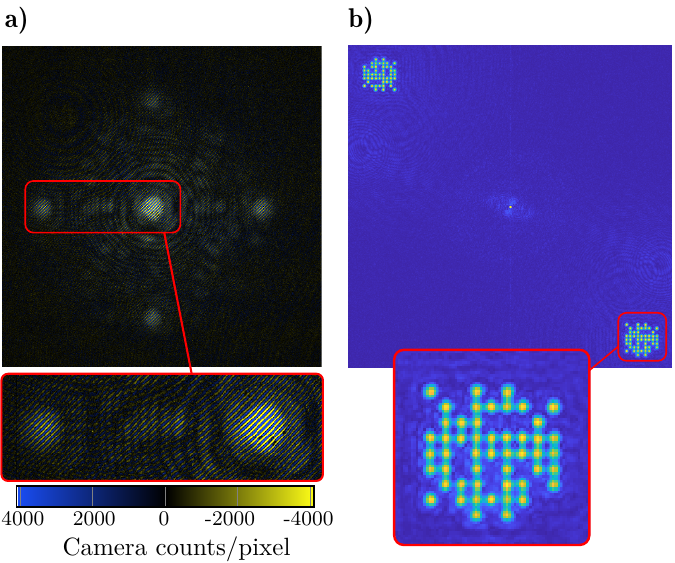}
\caption{a) Section of the recorded hologram $I_D- I_R$. For this recording, each hole of the mask scattered roughly 40,000 photons. 
	The panel below is a magnification. The color bar gives the number of counts per pixel. This count number  can be negative as we are dealing with a difference of two images. 
	 b) Section of the FFT of the hologram with a magnified view of the reconstructed hole pattern of the mask.  }
\label{fig:recorded_signal}
\end{figure}

Fig. \ref{fig:recorded_signal}a) depicts a hologram after $I_R$-subtraction, along with a magnified section. For this, the default mask shown in Fig. 2a) and an objective with a NA of 0.75 was used. The angles of the reference beam were $\theta = 0.84^{\circ}$, $\phi = 45^{\circ}$ \cite{note3}
and about 40,000 photons were transmitted through each hole of the mask.
 We only show the section of the hologram that contains  the relevant features. 
It exhibits five dominant spots, arranged in a cross-like fashion, with weaker signals in between.

The origin of the five dominant interference peaks can be understood as follows. 
To a first approximation, the holes in the mask form a 2D square lattice.
The far-field diffraction pattern of a 2D square lattice is again a square lattice.
The spot in the center of the hologram is the zeroth-order diffraction peak of this square lattice, while
the surrounding spots are  first-order peaks. 
 The array of holes in the mask, however, is not a perfect square lattice since a number of lattice sites are not occupied. As a consequence, the intensity in between the major diffraction peaks is non-zero and this is most relevant for the reconstruction of the hole positions.
The hologram is modulated  with high spatial frequency by a sinusoidal wave  at an angle $\phi = 45^\circ$.
This oscillatory pattern is due to the interference of the reference beam with the scattered probe beam.   
In the FFT it leads to a diagonal shift of the reconstructed  hole pattern of the mask from the center, see  
 Fig.\,\ref{fig:recorded_signal}b).  Mathematically this shift is equivalent to the shift of the
 vectors $\mathbf{k}_n$  by  $\mathbf{k}_R$, as previously discussed in the paragraph following Eq.\,(\ref{eq:k_n}). 
 As a result, the reconstructed hole pattern after the FFT is located in the upper left and lower right corners.
  The two patterns are inverted with respect to each other, as they correspond to opposite momenta 
 $\pm  (\mathbf{k}_n -  \mathbf{k}_R)$.
   
The shift of the
 reconstructed pattern is advantageous because it reduces noise.
 Without the shift, both patterns would be located in the center 
 where they would overlap with each other, with the noisy signal from the reference beam,  and with
  the $| E_\mathrm{S} |^2$ term in Eq.\,(\ref{eq:ID_SR1}). We find that a shift in diagonal direction is helpful because there the noise background is particularly small \cite{note3}.
    
  The FFT in Fig.~\ref{fig:recorded_signal}b) clearly reproduces  the hole pattern
  of mask a) in Fig.~\ref{fig:masks}, which shows that the holographic imaging scheme works.

    \begin{figure}
  	\includegraphics[width=0.48\textwidth]{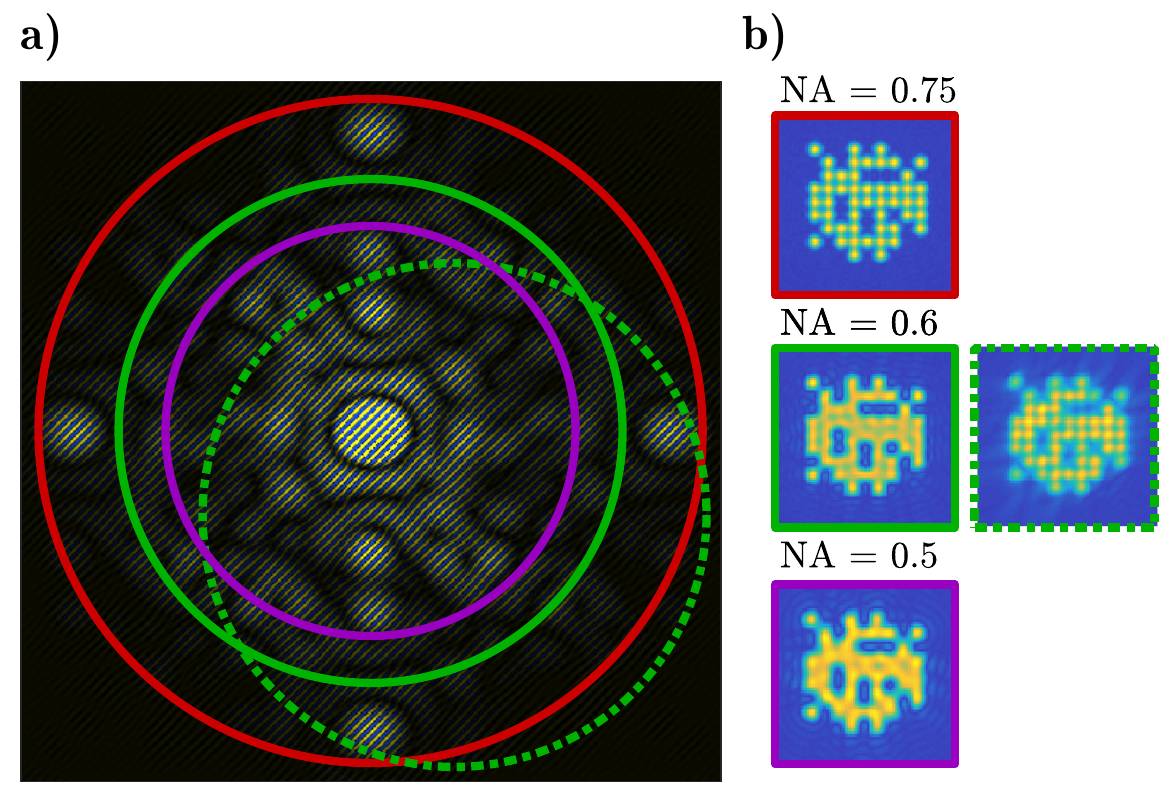}
  	\caption{ a) The circles drawn on the calculated diffraction pattern represent various numerical apertures: NA = 0.5 (purple), NA = 0.6 (green), and NA = 0.75 (red). For a given NA, only the pattern inside the circle ends up on the camera sensor.  
  	Continuous lines correspond to a hole mask that is centered on the optical axis of the lens, while for the green dashed line it is off-center.   	
  		b)  The relevant sections of the corresponding Fourier transforms are shown.}
  	\label{fig:NA}
  \end{figure}

 \section{Numerical aperture} \label{sec:NA}
  
  \vspace{0.1cm}
  According to Abbe's theory of imaging, the first order diffraction peaks
  of a lattice need to be recorded in order to clearly resolve the individual lattice sites. 
  Therefore, the numerical aperture (NA) of the microscope objective needs to be large enough. 
  Figure \,\ref{fig:NA} shows a calculated hologram for our default hole mask from Fig.~2a).
  If the hole mask is centered on the optical axis, the NA of the objective can be represented by a circle in  $\vec{k}$  momentum space. 
  In Fig.\,\ref{fig:NA}a)  such circles are drawn for NA = 0.5, 0.6, and 0.75. The relevant sections of the FFTs of the inner parts of the circles 
  are displayed in Fig.\,\ref{fig:NA}b). The sharpness of the hole pattern increases with increasing NA.

  When the mask is centered on the aperture of the objective with NA = 0.6 (green solid line), none of the first-order peaks are caught. By shifting the mask diagonally, however, one can include two first-order peaks  while still retaining the central area which includes most of the hologram's information (green dashed line). 
  (We note that for a small displacement of the mask, the solid angle at which light is collected by the objective decreases only minimally  and the resulting ellipse can be still approximated by a circle.)
  This inclusion of the first-order peaks can help to better resolve individual lattice sites. 
  However, because the positions of the sites of the regular lattice are known, 
   the hole pattern can still be clearly determined even for the centered case and low NA.

 \begin{figure}
 	\includegraphics[width=0.48\textwidth]{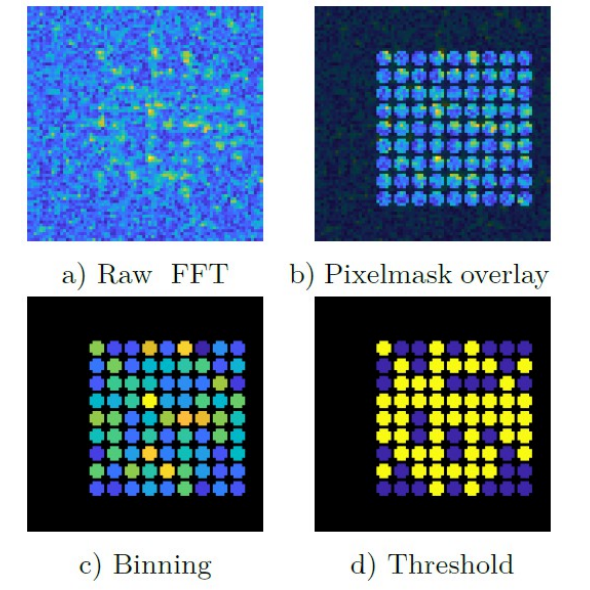}
 	\caption{Reconstruction of the mask for $N_{ph}=100$ photons per hole. a) Region of interest in the FFT. b) Processing via overlaying of a digital template. c) Binning of the pixels assigned to each lattice site.  d) Subsequent application of a threshold to distinguish between occupied and empty sites. For simplicity we chose here an experimental sample for which the hole pattern was correctly reproduced. For 100 photons we typically only assign 90$\%$ of the holes correctly, see Fig. \ref{fig:fidelity_mask}. }
 	\label{fig:reconstruction}
 \end{figure}

  \section{Photon shot noise} \label{sec:LowLight}
  
  We now investigate how the reconstruction quality of the hole pattern decreases as the 
  probe light power is lowered. 
  From Eq. (\ref{eq:ID_SR1}) it is clear that the holographic signal scales with the electrical field amplitude of the diffracted laser beam
  $E_S$ and therefore with the square root of the number of scattered photons per hole. 
  The noise, on the other hand,  is fundamentally dominated by the shot noise of the light of the reference beam,  corresponding to the $|E_R|^2$ term in  Eq. (\ref{eq:ID_SR1}).
    
  As a consequence, for a fixed value of the reference beam power, signal to noise diminishes for a lower probe beam power, or in other words, for a smaller number of scattered probe beam photons $N_{ph}$ per mask hole.
    
  We find that once $N_{ph}$ is reduced to below about 500, the signal to noise ratio is so weak that 
   a simple  determination by eye of the hole pattern is no longer possible.
  Fig.\,\ref{fig:reconstruction}a) shows the FFT image for an extreme case where the average photon number per hole was only  about 100.  With the following  algorithm we can still decide with high fidelity whether a lattice site is occupied or empty.
For this, we make use of the known positions of the lattice sites in the Fourier plane. 
A black pixelmask consisting of a 2D array of circular slots (see Fig.\,\ref{fig:reconstruction}b)) is overlaid 
with the FFT image such that the midpoints of the slots coincide with the positions of the lattice sites.
Within each slot the FFT signal is added up, yielding a value $S_n$ where the index $n$ labels the respective slot,    
see Fig.\,\ref{fig:reconstruction}c).
If the value $S_n$ of a lattice site is larger than an appropriate threshold value,
$S_n> S_{\rm thr}$, the site is declared to be  occupied, otherwise empty. 
The threshold value $S_{\rm thr}$ needs to be determined independently, e.g. by using a known hole pattern or by another statistical method such as \cite{note4}.\\
With this discrimination method, the assignment of the occupation of the lattice site becomes
a probabilistic process. We define the recognition fidelity $F$ as the probability that the assignment for the lattice site is correct.
\begin{figure}
	\includegraphics[width=0.48\textwidth]{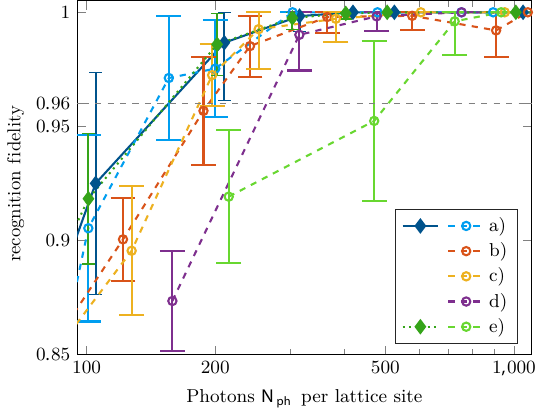}   
	\caption{Recognition fidelity $F$ for different masks (see Fig.\,\ref{fig:masks}) and photon numbers per hole.  Circles are experimental data. The settings for the measurements were NA = 0.75, $\theta = 0.64^{\circ}$, and $t_{exp} = 144\,\mu$s. 
		For each experimental data point we took $\sim$14 images  and determined the fidelity for each image. From this list of values the mean value and standard deviation were obtained. 	
		Diamonds are simulations which have been rescaled for better comparison with the experimental data. Namely, for a given calculated data point the actual photon number  $N_{ph}$ is 10 times smaller than indicated in the plot. For the simulations we use over 50 images per data point. Each image has a different (random) photon shot noise.
		The 96\% fidelity benchmark, which we chose arbitrarly, is represented by the black dashed line.}
	\label{fig:fidelity_mask}
\end{figure}

Figure\,\ref{fig:fidelity_mask} shows this fidelity $F$ as a function of $N_{ph}$ for the hole masks in 
Figs.\,\ref{fig:masks}\,a)\,-\,e).
Experimental data are shown as circles. 
From the diagram we infer that 300 
 diffracted photons per hole are sufficient to obtain a nearly perfect reconstruction of the hole arrays.

By lowering the probe beam intensity, the signal-to-noise ratio  
degrades and finally, below $N_{ph}\approx 300$,
the fidelity $F$ starts to decline. The characteristics of the decline is similar for all masks under study.

In addition to measuring  experimental fidelities, we also calculated them, using simulations as layed out in the  Appendix \ref{sec:simulations}.   While the calculations confirm the trend that the fidelity suddenly drops below a critical photon number, 
	the absolute agreement with the experiment is not good. For a given fidelity the calculated required photon number $N_{ph}$ is about a factor of
	10 smaller than for the experiment. In order to conveniently compare the trends of experiment and theory in Fig. \ref{fig:fidelity_mask}  we have rescaled the 
	theoretical $N_{ph}$ values, multiplying them by 10. These data are shown as diamonds. 
At this point it is not clear what the reason for the discrepancy between theory and experiment is. Possibly wavefront distortions of the light passing through optical lenses might play a role. This will be subject of future work.

\begin{figure*}
	\includegraphics[width=0.98\textwidth]{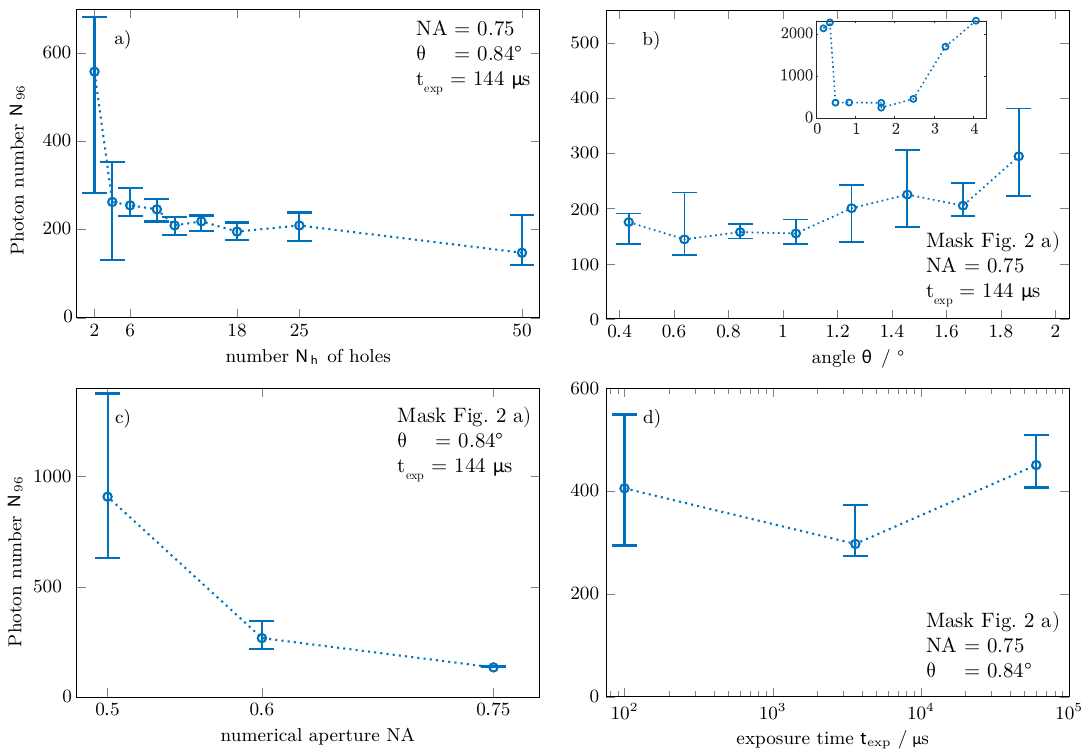}
	\caption{Required photon number $N_{96}$ per hole for 96\% recognition fidelity, plotted as a function of a) the number of holes in the mask, 
		b) the angle $\theta$ between reference beam and $z$-axis (see Fig.~\ref{fig:beams}), c) the numerical aperture NA of the microscope lens, and d) the camera exposure time (for constant total photon number). $N_{96}$ and the corresponding error bars are derived from an interpolation, see Fig. \ref{fig:fidelity_mask}.}
	\label{fig:N96}
\end{figure*}

In the following we investigate in  how far the onset of the decline depends
 on certain parameters of the set-up. This will provide us with the minimal number of photons that need
 to be scattered per hole to still achieve a high fidelity in the reconstruction of the hole pattern.

\section{Minimal photon number } \label{sec:}

In order to quantify the onset of the decline in fidelity,
we introduce the quantity $N_{96}$ which is the required number of photons per hole to achieve a fidelity of 96\%. 
It can be extracted from Fig.\,\ref{fig:fidelity_mask} by reading off the photon number $N_{ph}$ for which the 
data interpolations (colored lines) cross the  96\% fidelity line (gray dashed line).

In Figs.\,\ref{fig:N96}\,a)- d), $N_{96}$ is plotted as a function of four parameters. 
Figure\,\ref{fig:N96}\,a) shows $N_{96}$ as  a function of the number of holes $N_h$ in a mask.
A first glance at the measured data seems to indicate a decrease of $N_{96}$ with $N_h$. However, we note that given the error bars this decrease is statistically  not significant.

Figure\,\ref{fig:N96}\,b) shows that $N_{96}$ only moderately depends on the angle $\theta$ between
the reference beam and the $z$-axis within the range $0.4^\circ < \theta < 2^\circ$.
As already discussed in section  \ref{sec:recognition},
an angle $\theta$ that is too small leads to a reconstructed hole pattern which is overshadowed by noise in the vicinity of the center of the FFT. For a $\theta$ that is too large, the fringes in the hologram are too closely spaced and therefore cannot be resolved by the camera sensor. Using the Nyquist-Shannon sampling theorem we estimate that this limit sets in
at a critical angle of $\theta = 4^{\circ}$ for our experimental set-up. Therefore, if we reach angles that are either too small or too large, the experimentally determined numbers for $N_{96}$ strongly increase. This can be seen in the inset where we show a coarse scan from $\theta = 0.2^\circ$ to $4^\circ$.
For optimized settings in our experiment we chose the angle $\theta = 0.7^{\circ}$.

With  Fig.\,\ref{fig:N96}\,c) we return to our discussion in Sec.\,\ref{sec:NA} on how the reconstruction quality of the
hole pattern depends on the numerical aperture NA of the objective lens. 
For these data, the mask was centered on the optical axis of the microscope lens. 
We plot $N_{96}$ for NA = 0.75, 0.6, and 0.5. 
 The first order peaks in the hologram are only included for  NA = 0.75 (see also Fig. \ref{fig:NA}).
The experimental data show that  $N_{96}$ strongly increases as the NA is lowered.
This is in contradiction to our simulations in  Fig. \ref{fig:NA}b) where we found that that despite the blurring the the overall signal of a site did not strongly change.

Finally, in Fig.\,\ref{fig:N96}\,d) we study the dependence of $N_{96}$ on the exposure time of the digital camera. Here, the light intensity is adjusted such that the total numbers of photons from the probe and reference beams hitting the camera are kept constant. 
We do not observe a significant dependence on exposure time for the shown time window. This is expected as long as, e.g., mirror vibrations and long-term interferometric drifts, as well as accumulated thermal camera noise do not strongly affect the hologram.

\section{Summary and conclusion} \label{sec:conclusion}

We have successfully tested a recently proposed holographic method for imaging $\mu$m-scale patterns which are arranged on a 2D grid. Such patterns consist of a random array of submicron holes in an opaque mask.
 We experimentally and theoretically searched for the minimum number of photons that need to be scattered off the pattern in order to reconstruct the pattern holographically with high fidelity. 
 After optimization, we found experimentally that about 200 diffracted photons per hole are sufficient to reconstruct the hole positions in the masks with a fidelity of 96\%.	Our simulations predict that this number can still be improved by about a factor of 10. In the future we anticipate that this method can be applied 
 to image ultracold atoms in optical lattices with single-site and single-atom resolution, without the need of 
 additional cooling.

 \section{Acknowledgement}
 We would like to acknowledge financial support by the 
 German Research foundations (DFG) through grant\\ 382572300.
  The authors would like to thank Markus Deiss and Frederik Koschnick for proofreading the manuscript.

\section{Appendix} \label{sec:appendix}

\subsection{ Fabrication of the hole mask} \label{sec:holemasks}

Circular areas were exposed by means of a Leica EBPG 5 HR electron beam writer applied on fused silica photo mask blanks. 
The mask blanks (size: 100$\times$100\,mm$^2$, thickness: 2.3\,mm) were coated with chrome (thickness: 90\,nm, optical density: 3.0)  and a positive e-beam resist.
 After e-beam exposition and developing the round holes were produced by wet chemical etching. 
The finished structures were controlled by means of optical microscopy. Atomic force microscopy revealed a typical hole radius of 300$\pm$30\,nm. 
After fabrication the masks were protected with the polymer Crystalbond$^{\mathrm{TM}}$ and cut into square pieces ($\approx 25\times 25$\,mm$^2$). 

\subsection{ Properties of the digital camera}
\label{sec:digicam}

The CMOS sensor of the pco.edge 4.2LT camera has a pixel size of 6.5 $\mu$m $\times$ 6.5 $\mu$m. It has a digital resolution of 16 bit, 37,500:1 dynamic range, and 73\% quantum efficiency at 671\,nm. The full well depth is  about 30,000 electrons. Therefore, the signal saturates at about 40,000 photons/pixel.  
There is a signal conversion of 0.46 $e^-$/count. Dark current is negligible for our experiments. A short exposure without light has a constant offset of 100.3 $\pm 0.6$ counts and the corresponding  rms-noise is 2.2 counts. 
   The nominal readout noise is 1.3 $e^-$ (rms) which agrees roughly with the 2.2 count noise. 
The noise of our holographic signals is generally dominated by the photon shot noise. According to the Poisson distribution, if the average number of 
incoming photons is $N$, then the shot noise on that number is $\sqrt{N}$ (standard deviation).
Since the  conversion of photons into electrons is
probabilistic with probability  $p = 0.73$, the Poisson distribution for the photons is thinned out
to produce a Poisson distribution for the electrons with an expectation value 
(and variance) of $Np$, i.e. a shot noise of $\sqrt{Np}$.

\subsection{ Details of the simulation } \label{sec:simulations}

In the simulation shown in figure \ref{fig:fidelity_mask} the hole mask is represented by a matrix of square pixels, each with 160\,nm $\times$ 160\,nm size. A pixel which is located within a hole has a transmission of 1. A pixel which is located on the edge of a hole has a transmission lower than one, as only a part of pixel area is covered by the hole aperture. We assume the holes to be illuminated by a Gaussian probe laser field. We calculate the FFT of the electrical field amplitude of the transmitted light and clip off parts which lie outside the numerical aperture of the lens. This results in the electrical field amplitude of the probe laser at the plane of the CCD sensor.  This field is superposed with the electrical field amplitude of the Gaussian beam of the reference laser. We take into account signal loss due to the finite quantum efficiency of the camera, the finite transmission of the NPBS, and reflections on optical surfaces.  Next, we calculate the expectation value of the photon count for each pixel on the CCD chip and add photon shot noise. Photon shot noise strongly dominates over other noise sources such as  the read-out and thermal noise of the CMOS camera and speckle noise. Speckle noise takes into account interference fringes originating from dust particles on the optics and from apertures and we use a speckle noise model as described in \cite{Hof16}.


\begin{thebibliography}{12}



\bibitem{Bak09}
W. S. Bakr,  J. I. Gillen, A. Peng, S. F\"olling,  and M. Greiner,
\newblock Nature \textbf{462}, 74 (2009).

\bibitem{She10}
J. Sherson, C. Weitenberg, M. Endres,	M. Cheneau, I. Bloch,  and S. Kuhr,
\newblock Nature \textbf{467}, 68--72 (2010).


\bibitem{Che15}
L. Cheuk, M. Nichols, M. Okan, T. Gersdorf, V. Ramasesh, W. Bakr, T. Lompe, and M. Zwierlein,
\newblock Phys. Rev. Lett. \textbf{114}, 193001 (2015).

\bibitem{Hal15}
E. Haller, J. Hudson, A. Kelly, D. Cotta, B. Peaudecerf, G. Bruce, and S. Kuhr,
\newblock Nature Phys. \textbf{11}, 738 (2015).

\bibitem{Par15}
M. Parsons, F. Huber, A. Mazurenko, C. Chiu, W. Setiawan, K. Wooley-Brown, S. Blatt, and M. Greiner,
\newblock Phys. Rev. Lett. \textbf{114}, 213002 (2015).

\bibitem {Gro15}
A. Omran, M. Boll, T. Hilker, K. Kleinlein, G. Salomon, I. Bloch, and C. Gross,
\newblock Phys. Rev. Lett. \textbf{115}, 263001 (2015).


\bibitem{Ber18}
A. Bergschneider, V. M. Klinkhamer, J. H. Becher, R. Klemt, G. Zürn, P. M. Preiss, and S. Jochim,
\newblock Phys. Rev. A \textbf{97}, 063613 (2018).


\bibitem{Ger08}
T. Gericke, P. Würtz, D. Reitz, T. Langen, and H. Ott,
Nat. Phys. \textbf{4}, 949 (2008).

\bibitem{Vei21}
C. Veit, N. Zuber, O. A. Herrera-Sancho, V. S. V. Anasuri, T. Schmid, F. Meinert, R. Löw, and T. Pfau,
\newblock Phys. Rev. X \textbf{11}, 011036 (2021).

\bibitem{Ott16}
H. Ott,
\newblock Rep. Prog. Phys. \textbf{79}, 054401 (2016).


\bibitem{Kad01}
S. Kadlecek, J. Sebby, R. Newell, and T. Walker,
\newblock Opt. Lett. \textbf{26}, 137-139 (2001).

\bibitem{Tur05}
L. Turner, K. Domen, and R. Scholten,
\newblock Phys. Rev. A \textbf{72}, 031403(R) (2005).

\bibitem{Wu14}
J. Sobol, and S. Wu,
\newblock New. J. Phys. \textbf{16}, 093064 (2014).



\bibitem{note1}
We note that many optical lattice set-ups in the literature exhibit lattice constants of about 500\,nm. 
Performing holography with a smaller lattice constant as compared to ours requires an optical lens with a correspondingly larger NA than ours.

\bibitem{note2}
In order {\it not} to average out the interference fringes between probe beam and reference beam, the averaging is done in a diagonal fashion. This is because the interference fringes for the given hologram are at an angle of 45$^\circ$ (see also Fig. \ref{fig:recorded_signal}a), as set by the chosen angle $\theta$ of the reference beam. Concretely, the diagonal averaging is calculated as $ \sum_{i=-5}^{5} I_{D,R}(x+i, +i  ) / 11 $ over 11 pixel rows.

\bibitem{note3}
We choose $\theta$ = 45$^\circ$ in the experiment because this has technical advantages. For one, diagonal pixel lines have a distance which is reduced by a factor of $\sqrt{2} $ as compared to the horizontal and vertical pixel lines. This increases the spatial resolution for measuring fringes. Second, the angle of $\theta$ = 45$^\circ$ shifts the holographic signal away from spurious horizontal and vertical lines running through the center of the hologram. These lines  stem from clipping of the reference beam at the edges of the camera chip. Figure \ref{fig:recorded_signal}b) shows such line which is weak and  runs vertically through the center of the hologram.  
\bibitem{note4}
For a variety of lattice sites many measurements of the occupation signals are taken and a histogram of the occupation signals is generated. Ideally, the histogram will exhibit two peaks, corresponding to an empty and occupied site. The minimum between the two peaks can then be used to set the threshold value $S_{\rm thr}$, see also \cite{Hof16}.

\bibitem{Hof16}
D.K. Hoffmann, B. Deissler, W. Limmer, and J. Hecker Denschlag,
\newblock Appl. Phys. B \textbf{122},  227 (2016).

\bibitem{Kim11}
M.K. Kim, \textit{Digital Holographic Microscopy - Principles, Techniques, and Applications} (Springer, New York 2011).

\bibitem{Ver15}
N. Verrier, D. Donnarumma, G. Tessier, M. Gross,
\newblock Appl. Opt. \textbf{54}, 9540 (2015).


\end{thebibliography}
\end{document}